\documentclass[aps,amsmath,amssymb,prl,superscriptaddress,floatfix,twocolumn,notitlepage]{revtex4-1}

\usepackage{bm}
\usepackage{epsfig}
\usepackage[usenames]{color}
\usepackage{graphicx}
\usepackage{amsfonts}

\usepackage{cyrillic} 
\newcommand{\cyrrm}{\fontencoding{OT2}\selectfont\textcyrup}

\newcommand{\nn}{\nonumber}

\renewcommand{\paragraph}[1]{\textit{#1.---} }

\DeclareMathOperator{\Imag}{Im}

\newcommand{\junc}[1]{^{(#1)}}  
\newcommand{\grain}[1]{_{#1}}

\begin{document}

\title{Localization-Delocalization Transition and Current Fractalization}

\author{N.~M.~Chtchelkatchev}
\affiliation{Materials Science Division, Argonne National Laboratory, Argonne, Illinois 60439, USA}
\affiliation{Institute for High Pressure Physics, Russian Academy of Science, Troitsk 142190, Russia}
\affiliation{Department of Theoretical Physics, Moscow Institute of Physics and Technology, 141700 Moscow, Russia}

\author{A.~Petkovi\'c}
\affiliation{Laboratoire de Physique Theorique-CNRS, Ecole Normale Superieure, 75005 Paris, France}
\affiliation{Materials Science Division, Argonne National Laboratory, Argonne, Illinois 60439, USA}

\author{A.~Glatz}
\affiliation{Materials Science Division, Argonne National Laboratory, Argonne, Illinois 60439, USA}

\author{T.\,I.\,Baturina}
\affiliation{Materials Science Division, Argonne National Laboratory, Argonne, Illinois 60439, USA}
\affiliation{A. V. Rzhanov Institute of Semiconductor Physics SB RAS, 13 Lavrentjev Avenue, 
Novosibirsk, 630090 Russia}

\author{V.~M.~Vinokur}
\affiliation{Materials Science Division, Argonne National Laboratory, Argonne, Illinois 60439, USA}

\date{\today}

\begin{abstract}
We develop an analytical theory of the localization-delocalization transition for a disordered Bose system,
focusing on a Cooper-pair insulator.
We consider a chain of small superconducting granules coupled
via Josephson links and show that the
low-temperature tunnelling transport of Cooper pairs is mediated by a self-generated environment
of dipole excitations comprised of the same particles as the tunnelling charge carriers in accord with the early notion
by Fleishman, Licciardello, and Anderson~\cite{Anderson1978}.
We derive an analytical expression for the current-voltage characteristic
and find that at temperatures, $T$, below the the charging energy of a single junction, $E_c$, 
the dc transport is completely locked by Coulomb blockade effect at all voltages except for a discrete set of resonant ones.
At $T>E_c$ the combined action of disorder and temperature unlocks
the charge transport, since the environment excitation spectrum becomes quasi-continuous according to a
Landau-Hopf-like~\cite{Landau,Hopf,Landau-Hydrodynamics,noteLH} scenario of turbulence, and the conductivity acquires an
Arrhenius-like thermal activation form.  The transition from the localized to delocalized
behaviour occurs at $T=E_c$ which corresponds to the onset of turbulence in the
spectral flow of environmental excitations with Reynolds number ${\cal R}e\equiv(k_{\scriptscriptstyle B}T/E_c)=1$.
The proposed theory breaks ground for a quantitative description of dynamic and quantum phase transitions in a wealth of physical systems ranging from cold atoms in optical lattices, through disordered films and wires to granular and nanopatterned materials.
\end{abstract}



\maketitle

\section{Introduction}

In strongly disordered systems the low-temperature charge transport occurs via tunnelling between localized
states, having in general  different energy levels.
Thus tunnelling is possible only if charge carriers can either emit the excess or absorb the deficit in energy to accommodate
the difference between the initial and final states [see inset in Fig.~\ref{fig.ring}a)].  This process is referred to as the energy relaxation to a
\textit{bosonic environment}.  The most common environment is a phonon bath;
recent studies revealed, however~\cite{Gornyi2005,Lopatin2005,BAA2006,BAA,Chtch2009} that
in Josephson-junction arrays (JJAs) and granular materials another relaxation mechanism naturally arises: emission/absorption
of dipole excitations (charge-`anti-charge' pairs) comprised of the same particles that carry the current.

This self-generated dipole environment possesses an infinite number of degrees of freedom and thus
serves as the thermostat itself, dominating over the usual relaxation mechanism via phonons, which becomes inefficient
at low temperatures.
Indeed, in highly disordered
materials the standard electron-phonon interaction via a deformation potential should produce
phonons with wavelength on the order of the electron wavelength
$\lambda$, which, in its turn, is on the order of the localization length. In granular materials or JJAs the role of the localization length
is taken by the granule size (superconducting granules in the case of a JJA). Thus the characteristic phonon energy is
 $E_{ph}\simeq \hbar q s\sim \hbar s/\lambda $,
and the corresponding phonon temperature is $T_{ph}=E_{ph}/k_{\scriptscriptstyle B}$ which for typical localization lengths on the
order of 10\,nm is about 10\,K.  At temperatures $T\ll T_{ph}$ relaxation via phonons becomes inefficient
and gives way to \textit{non-phonon} mechanisms.

Since phonons are not involved into the process of  tunnel charge transfer, the related \textit{heating processes}
in JJAs and granular systems, which can be referred to as electronic insulators, can be expected significantly reduced as compared to conventional insulators.
The suppression of heating when relaxation occurs via energy exchange with a non-phonon environment
was indeed found in single-junction systems~\cite{GCBV}.

To proceed to a theory of bosonic insulator we consider an exemplary system, a one-dimensional (1D) JJA,
 a tunable and experimentally accessible realization of a generic
1D strongly interacting (charged) disordered boson chain,  modelling
 a wealth of physical systems ranging from helium in vycor~\cite{Crowell1995}, localized Cooper pairs near the
superconductor-insulator transitions in thin film~\cite{Shahar2005,Baturina2007}, and ultracold atoms~\cite{Damski2003,Lye2007,Luhmann2008,White2009}.
At the same time, in the limit where the capacitance of a single junction well exceeds its capacitance to the ground,
the 1D JJA is equivalent to a chain of coupled quantum rotors and
 allows for an analytical description. This makes it
a unique theoretical laboratory for studying disorder-induced non-conventional insulators with physics arising from
the  interplay between strong disorder, quantum fluctuations, and strong Coulomb interactions.


\section{Model and tunnelling transport in one-dimensional Josephson junctions array}

The behaviour of JJAs is controlled by two competing parameters, $\bar E_c=\sum E_c\junc{i}/N$,
the average charging energy of a single superconducting junction and $E_J$, the average Josephson coupling energy between
neighbouring superconducting granules. Here $N$ is the length of the array and $E_c\junc{i}=e^2/C\junc{i}$ the charging energies of the individual junctions defined by their capacitances, $C\junc{i}$.  If $\bar E_c<E_J$ the system is
superconducting and $\bar E_c>E_J$ corresponds to an insulating state.
We focus on a deep insulating state, $\bar E_c\gg E_J$, treating $E_J$ as a perturbation.
At the same we let the disorder-induced dispersion in $E_c\junc{i}$ be also large as compared to $E_J$, such that
a continuous conduction band due to overlapping localized wave functions does not form.
We start with a qualitative picture of Cooper pair transport, that can be constructed in the context of simple system consisting of a single superconducting granule confined between the two electrodes.

The tunnel current is controlled by the intensity of the relaxation  and assumes the form~\cite{GrabertDevoret}
	\begin{equation}\label{simplecurrent}
	     I\propto\exp(-E/W)\,,
	\end{equation}
where $E$ is some characteristic energy associated with the tunnelling process and $W$ quantifies the relaxation rate.
One can expect $W$ to be proportional to the density of environmental excitations.
At high temperatures (exceeding the local gap
in the excitations spectrum) the equipartition theorem yields $W\simeq T$.
At low temperatures, where the discreteness of the
environment spectrum becomes essential, the relaxation rate is strongly suppressed and the tunnelling current is locked.

\begin{figure}[thb]
 \includegraphics[width=0.9\columnwidth]{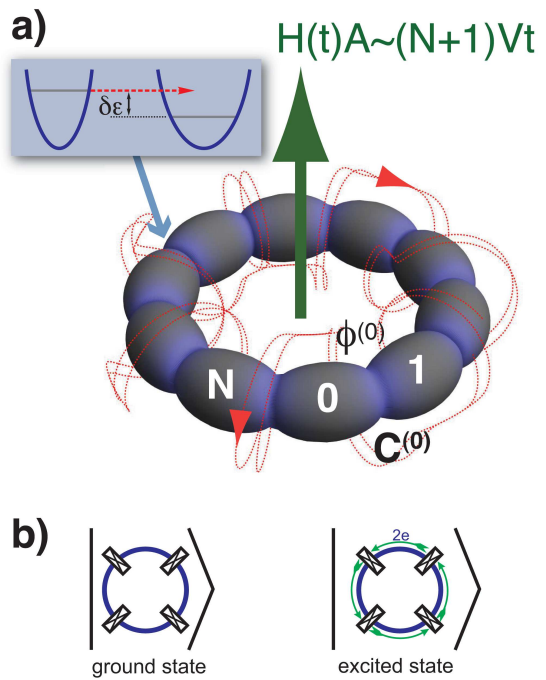}\\
  \caption{{\bf Model system} a) A circular array of Josephson junctions containing $N+1$ junctions (superconductors are dark gray, tunnel barriers blue), where $\phi\junc{i}$ and $C\junc{i}$ are the phase difference and the capacitance across the junctions between grains $i$ and $i+1$, respectively. The current in the array is induced by a time dependent perpendicular magnetic field $H(t)$. The total voltage follows from the time derivative of the total flux $\Phi(t)=H(t)A$ ($A$ denotes the area of the ring). The inset shows a sketch of a tunnelling process between two mesoscopic potential well.  Since the energy levels in these wells are in general different (due to
structural disorder), quantum mechanical tunnelling is only possible if the tunnelling particle emits (or absorbs) bosonic excitation which
can accommodate the energy difference $\delta\varepsilon$ between the respective energy levels.  b)
Illustration of {\it Left}: ground state and {\it Right}: exicted current states in the JJA ring}\label{fig.ring}
\end{figure}

Although this simple line of reasoning does not apply straightforwardly and neither is a single quantity $W$ characterizing the relaxation rate well defined~\cite{Chtch2009}, it gives a good
qualitative idea of the tunnelling transport in insulators and the nature of the localization-delocalization transition.
 To avoid the
problem of boundary conditions at the current leads we  close JJA into a ring,
the current being induced by a time changing magnetic field perpendicular
to the ring plane, see Fig.~\ref{fig.ring}a).
The current flowing through the series of junctions is the same
along the ring and one can therefore calculate it across an arbitrary junction
(i.e. between an arbitrary pair of adjacent granules).
The current is determined by the correlation function of superconducting phases along
the ring, describing the probability of the energy exchange between the tunnelling Cooper pair and the environment.
The correlation function is composed of matrix elements between the states where the number of
Cooper pairs at a given granule differs by one, i.e. the matrix elements between the states $|n\junc{i}\rangle$ and $|n\junc{i+1}\rangle$ ($n\junc{i}$ is the number of  Cooper pairs at the $i$-ths junction between superconductor $i$ and $i+1$).

Switching between these states means exciting a current state in the ring, see Fig.~\ref{fig.ring}b),
i.e. removing a Cooper pair from the $i$-ths granule and adding it to the $i+1$-ths granule and so on eventually
encircling the system, $i\to i-1\to i-2 ... \to i+2\to i+1$.
In the insulating limit the current is due to tunnelling and local Coulomb blockade effects
make the local environmental spectrum discrete (with a minigap of the order of the charging energy of a single junction $\bar E_c$),
which means that environmental dipole excitations get localized.
Localization of environmental excitations suppresses the relaxation and thus completely locks the current.
However, as we show below the combined effect of disorder and temperature
gives rise to Landau-Hopf-like turbulence in the spectral flow of the environment levels as soon as the ratio $T/\bar E_c$, which plays the role of the Reynolds
number in our system, becomes large.  As a result the spectrum at temperatures above the minigap becomes locally quasi-continuous unlocking the current.
At $T\ll \bar E_c$ the spectral flow of the environmental levels can be characterized as rather ``laminar," the local spectrum
retains its discrete nature and tunnelling current remains significantly suppressed.

The JJA loop is described by the quantum rotor Lagrangian:
\begin{eqnarray}
\mathcal{L}&=&\mathcal{L}_C+\mathcal{L}_J,\\
\mathcal{L}_C&=&\frac{1}{8e^2}\sum_{i}C^{(i)}(\dot{\chi}_{i}-\dot{\chi}_{i+1})^2 +\frac{1}{8e^2}\sum_{i}C_{0,i}\dot{\chi}_{i}^2,\\
\mathcal{L}_J&=& - \sum_{i}E_J^{(i)}\cos\left(\chi_{i}-\chi_{i+1}\right)\,,
\end{eqnarray}
where $\chi\grain{i}$ is the phase of the order parameter at the superconductor $i$, $\chi\grain{N+1}=\chi\grain{0}$,
$C\junc{i}$ and $E_J\junc{i}$ are the capacitance and the Josephson
coupling of the junction connecting the $i$-ths and $(i+1)$-ths grains, respectively - 
we use the shorthand notation $(i)$ for junctions $(i,i+1)$.
Quantities related to junctions use the superscript $(i)$ and subscripts $i$ label single grain properties.
The terms $\mathcal{L}_C$ and $\mathcal{L}_J$ describe the charging energy 
and the Josephson coupling between adjacent superconducting islands, respectively.
The capacitance to the ground of the $i$-ths superconducting grain is $C_{0,i}$.
Here the convention $\hbar=c=k_B=1$ is used and lengths are measured in units of a single junction size.
We consider the deep insulating state where 
$E_c\junc{i}=e^2/C\junc{i}$, $E_{c_{0,i}}=e^2/C_{0,i}\gg E_J\junc{i}$.
The bias $V$ is the voltage drop across the single junction. 
We address the most common experimental situation where $\max C_{0,i}\ll\min C\junc{i}$ and
focus on the case where the charge screening length $\Lambda=(\bar C/\bar C_0)^{1/2}>N$, 
the generalization onto the opposite limit is straightforward. The bar denotes averaging over grains.

\begin{figure}[thb]
  \includegraphics[width=0.9\columnwidth]{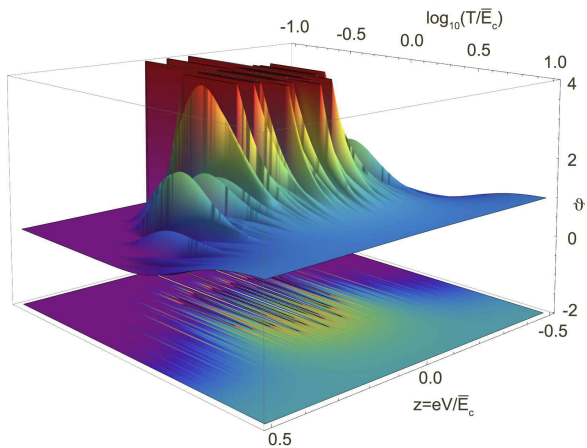}\\
  \caption{{\bf Matrix-Theta function} Semi-logarithmic surface plot of the matrix $\vartheta$-function which defines the current in the systems as function of voltage and temperature fro $N=8$, $\sigma=10^{-4}\bar{E}_c$, and charging energy variance $\alpha=0.056$.}\label{fig_oscillations}
\end{figure}

Treating ${\cal L}_J$ as a perturbation~\cite{GrabertDevoret}, one finds the supercurrent in junction $(i)$:
      \begin{align} \label{eq:Is}
               I_s(V\junc{i})=&-4e[E_{\scriptscriptstyle J}^{(i)}]^2\int dt \sin \left[ 2eV\junc{i}t\right] \Imag [\mathcal{K}(t)].
         \end{align}
Here $V\junc{i}$ is the voltage drop at the junction between grain $i$ and $i+1$ and
          \begin{gather}\label{eq:K}
                  \mathcal{K}(t)=\langle \exp[\imath \phi\junc{i}(t)]\exp[-\imath \phi\junc{i}(0)]\rangle_{{\cal H}_C},
           \end{gather}
where $ \phi\junc{i}(t)$ is the phase difference across the junction, describing the fluctuations around the mean value determined by the external voltage. Using equations~\eqref{eq:Is} and \eqref{eq:K} one can derive an analytical description of
Cooper pair tunnelling transport in 1D JJAs.
The Fourier transform of the  phase correlation function, $\mathcal{K}(\omega)$,
		\begin{eqnarray}
				 \mathcal{K}(\omega)&=&2\pi\sum_{\{n\},\{m\}}P_{\{n\}}\left|\langle{\{n\}}|e^{i\phi\junc{i}}| {\{m\}}\rangle\right|^2\times\nn\\
				 &&\delta(E_{\{n\}}-E_{\{m\}}+\omega),\label{eq.Komega.M}
		\end{eqnarray}
where $| {\{n\}}\rangle$ and $E_{\{n\}}$ denote the wave functions and energy levels for configuration $\{ n\}$ of $\cal{L}_C$, accordingly.
The Gibbs weight is given by $P_{\{n\}}=Z^{-1}{e^{-\frac{E_{\{n\}}}{T}}}$, with $Z=\sqrt{(2\pi T/\bar E_c)^{N}/\prod e\junc{i}}\sum_{\{m\}}e^{-2\pi^2 T  [m\junc{i}]^2 /e\junc{i}\bar E_c}$, $e\junc{i}=E\junc{i}_c/\bar E_c$ [this expression for $Z$ is obtained by using Poisson's summation formula].
$\mathcal{K}(\omega)$ has the physical meaning of the probability of exchanging an energy quanta
 $\omega$ with the environment, which is comprised of the excitations of the system itself.
This means, the system generates its own environment.
The charging energy of a junction $(i)$ is equal to $\epsilon\junc{i}(n\junc{i})=2E_c\junc{i} [n\junc{i}]^2$,
where the quantum numbers $n\junc{i}=0,\pm 1,\ldots$, $i=0,\ldots, N$ have a physical meaning of an excess (or deficit)
number of Cooper pairs at the junction.
Therefore  $E_{\{n\}}=\sum\epsilon\junc{i}(n\junc{i})$, where $\{n\}=(n\junc{0},n\junc{1},\ldots)$.
The wave function describing the quantum states of the annular JJAs is
		\begin{equation}\label{wavefunction}
			| {\{n\}}\rangle=\prod_{i=0}^N \psi_{n\junc{i}}(\phi\junc{i}),
		\end{equation}
where $\sum_{i=0}^N\phi\junc{i}=0$ and $\psi_{n}(\phi)=\exp\{\imath n\phi\}/\sqrt{2\pi}$ is the wave function of the quantum rotor.
The matrix element of the operator $\exp[-\imath \phi\junc{i}(0)]$ in \eqref{eq:K} is
nonzero for transitions where either (a) only $n\junc{i}$ changes by one
or (b) where $n\junc{i}$ is fixed but the quantum numbers of the rest of the junctions change by one simultaneously.
Integrating the most singular contributions to the current explicitly 
one arrives at
		\begin{gather}\label{eq:Is_2n_gg_ap}
  			  I_s= \left\{e^{-\frac{(eV+ N\bar E_c)^2}{2TN\bar E_c}}-
  			  e^{-\frac{(eV-N\bar E_c)^2}{2TN\bar E_c}}\right\} \vartheta\left({eV}/{\bar E_c},\mathcal T\right)\mathcal I,
		\end{gather}
where $\mathcal I=\frac{e\pi E_{\scriptscriptstyle J}^2\sqrt{ (2\pi T)^{N-1}}}{Z  \sqrt{ N\bar E_c^{N+1}\prod_i e_i}}$,
 $e_i=E_c^{(i)}/\bar{E}_c$, $Z=\sum_{\{m\}} e^{-2 E_c^{(i)}m_i^2/T}$, $z= V/\bar E_c$.
In the partition function $Z$, the sum is taken over all possible combinations of 
integer numbers $\{m\}=(m_1,m_2\ldots m_N)$. 
The generalized
 Jacobi  $\vartheta$-functions is defined as~\cite{Abramoviz-Stegun}:
		\begin{gather}\label{eq:theta}
    			\vartheta(z,\mathcal T)=\sum_{\{m\}} e^{\imath 2\pi z \vec m\cdot \vec a -\pi \vec m^\tau\mathcal T\vec m}\,,
		\end{gather}
where $\vec a=(1,1,\ldots,1)^\tau/N$ and the sum is taken over all integer vectors $\vec m=(m_1,m_2,\ldots,m_N)$. The matrix $\mathcal T$ is
parametrized via the vector $\vec e=(e\junc{1},e\junc{2},\ldots e\junc{N})^\tau$:
		\begin{gather}\notag 
    			\mathcal T_{ij}=\frac{2\pi T}{\bar E_c} H_{ij}+\frac{\pi\sigma^2}{2(N\bar E_c)^2}\,,
    			\qquad
   			 H_{ij}=\left(\frac{1}{ e\junc{i}}\delta_{ij}-\frac1N\right).
		\end{gather}
The quantity $\sigma\ll \bar{E}_c$ introduces the finite width of the quantum levels.  
The matrix $\mathcal T$ is positively defined, since the matrix $H$ has 
one zero eigenvalue corresponding to the eigenvector
$\vec h^{(0)}=\vec e$ while other eigenvalues of $H$ are of the order of unity.
%
%

We parametrize the dispersion in charging energy as 
$\delta E_c\equiv(\sum(E_c^{(i)}-\bar E_c)^2/N)^{1/2}=\alpha \bar E_c$. 
In the ``clean'' limit, $\alpha=0$, the theta-function structure is trivial: $\vartheta(z,\mathcal T)\propto \sum_n\delta(z+n)$.
For finite disorder, i.e. $\alpha>0$, the density of the $\delta$-functions is changed making $\vartheta(z,\mathcal T)$ ``nearly continuous'' at $T\gtrsim \bar E_c$.
The set of $\vec m$ in~\eqref{eq:theta} at high temperatures becomes restricted, because only the $\vec m$-configurations
directed close to $\vec h^{(0)}$ contribute to the $\vartheta$-function. 
At high temperatures, $T>\bar{E}_c$, 
Eq.~\eqref{eq:theta} reduces to
	\begin{gather}\label{eq:Is_2n_g_apa111}
    		\vartheta(z,\tau)=\sum_{n=-\infty}^\infty e^{\imath 2\pi n z-\pi n^2 \tau}\,,
	\end{gather}
where $\tau=2\pi\left[\frac{\sigma^2}{4\bar E_c^2}+ \frac{T\alpha^2 N} {\bar E_c}\right]$ for $\alpha< 1$,
and $\mathcal I\approx eE_j^2\sqrt{{\pi }/({2N T \overline{E}_c})}$. If $\tau>1$ then $\vartheta(z)\approx 1$ and if $\tau\leq 1$ then the $\vartheta(z)$ oscillates with the period $1$. While $\sigma\ll\bar E_c$, $T\alpha^2 N/\bar E_c$
can be of the order of unity even at weak disorder for $T> \bar E_c$.  
Thus, it follows from Eq.~\eqref{eq:Is_2n_g_apa111} that the disorder-induced term in $\tau$ 
dominates and the combined action of disorder and temperature makes the current a
smooth function of voltage.  Note that Eq.\eqref{eq:Is_2n_g_apa111} holds provided
$\sigma\gtrsim\alpha \bar E_c/ N!$.  This condition is not restrictive in large systems
and in the thermodynamic limit even infinitesimal broadening of quantum levels 
regularizes $I(V)$, which at high temperatures
does not depend on the particular choice of $\sigma$ because a chaotic fractalized
(turbulent) state develops.

The current-voltage characteristics of the JJA is governed by the analytical properties of the
$\vartheta$-function.  The behaviour of the $\vartheta$-function as a function of the
dimensionless voltage and temperature (measured in the units of ${\bar E}_c$) is illustrated in Fig.~\ref{fig_oscillations}.
One sees that the $\vartheta$-function is smooth at high temperatures but transforms into a set of singular peaks
at resonant voltages upon lowering the temperature.  
In other words, the $\vartheta$-function offers an analytical description
of the finite-temperature phase transition between conducting and localized states.

\section{Fractalization of the environmental spectrum and unlocked tunnel transport}

The dipoles constituting the environment are comprised of excessive Cooper pairs and Cooper ``holes" with charges
$-2e$ ($+2e$) and have the characteristic energy ${\bar E}_c$. If all the junctions in the JJA were identical, the environment energy spectrum would have consisted of
narrow bands $\sim N E_J<E_c$ separated by gaps $ E_c$.
Therefore, the current would have flown only at resonant voltages $V\simeq n{ E}_c/e$, where $n$ is integer.
If the charging energies are different (disorder in the charging energies also accounts for random 
offset charges that appear in the substrate),
the tunnelling Cooper pair can  relax the energy $\sum_i n_i E_c\junc{i}$ to the environment, with $n_i$ integer~\cite{relax}.
If the charging energies are incommensurate, the latter sum can become arbitrarily small.
One can further show that a broadening of quantum environmental levels to width $\sigma$, such that
$\alpha {\bar E}_c<\sigma N!$, where $\alpha{\bar E}_c$ is the dispersion of the charging energies,
turns the local environment excitations spectrum effectively quasi-continuous at temperatures $T>{\bar E}_c$.
At these temperatures the current-voltage ($I$-$V$) characteristics is well defined and smooth.
At $T>{\bar E}_c/(\alpha^2 N)$ the Coulomb blockade modulations in the $I$-$V$ curve almost vanish.
If the system is not very large such that (for $T>\bar{E}_c$) $T \alpha^2 N<\bar{E}_c$,
the modulations of the $I$-$V$ curves appear at voltages that are integer multiples of 
${\bar E}_c$, and near $T\simeq {\bar E}_c$,
the voltage scale of current modulations becomes of the order of $\alpha {\bar E}_c$.

\begin{figure*}[thb]
   \includegraphics[width=0.9\textwidth]{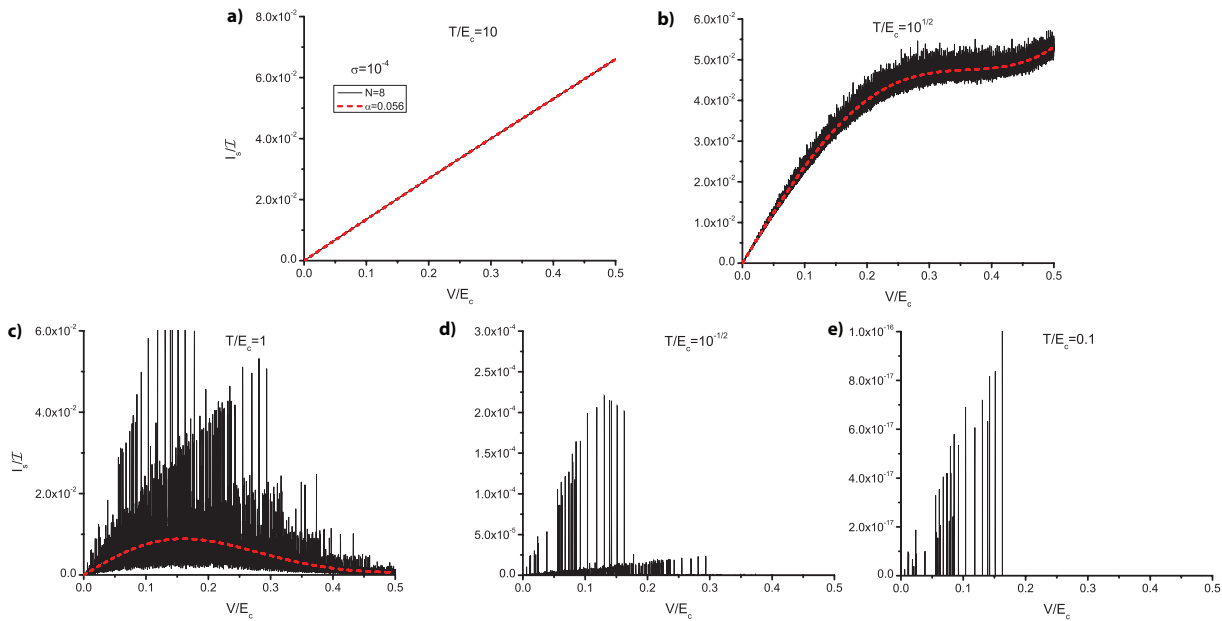}\\
  \caption{{\bf Current fractalization} $I$-$V$ characteristics at temperatures $T/\bar{E}_c=10^{n/2}$ [$n=2,\ldots,-2$, panels a)-e)] according to Eq.~(\ref{eq:Is_2n_gg_ap}). At larger temperatures the $I$-$V$characteristics is approximated by the scalar theta function with $\alpha=0.056$ (red dashed line). For the highest temperature ($n=2$), the slope is given by $2eV/Te^{-N\bar{E}_c/2T}$ ($\sim 0.13 eV/\bar{E}_c$ for the used parameters).
 All plots are for $N=8$ and $\sigma=10^{-4}\bar{E}_c$.
   }\label{fig.IV}
\end{figure*}

A quantitative description of fractalization of the environment energy spectrum and the 
smoothing of Coulomb blockade effects
can be achieved by expressing the correlation function as
\begin{align}\label{eq:Kt}
    			\mathcal{K}(t)&\equiv\langle \exp[\imath \phi\junc{i}(t)]\exp[-\imath \phi\junc{i}(0)]\rangle\\\label{eq:turbulence}&=e^{\imath\Omega t}\sum A_{p_1p_2\ldots p_N} \exp\left\{\imath\sum_{i=1}^N p_i\,\varphi\junc{i}(t)\right\}\,,
\end{align}
which is a measure for the temporal evolution of the phase determining the current via $I_s\propto\int dt\,\sin(2eVt)\Imag{\cal K}(t)$. The average is done with respect to the charging energy part of the Hamiltonian. Here $\varphi\junc{i}=4E_c^{(i)}t$, and the sum is taken over integers $p_1,\ldots,p_N$, $\Omega=-2N\bar E_c$, and
$A_{p_1p_2\ldots p_N}=Z^{-1}e^{-\sum_i2E_c\junc{i} p_i^2/T}$.
Noticing that Eq.~(\ref{eq:turbulence}) in fluid dynamics describes the velocity field in the 
liquid in which a turbulent flow develops,
one realizes that the time evolution of $\mathcal{K}(t)$ is governed by the Landau-Hopf scenario of turbulence~\cite{Landau,Hopf,Kolmogorov,Turbulence} with  Reynolds number ${\cal R}e\equiv T/\bar{E}_c$.
A sequence of incommensurate frequencies is generated by bifurcations occurring at ${\cal R}e=1$.

The effective level broadening in the high-temperature regime, $T>{\bar E}_c$,
is then $\delta E=2\pi\bar{E}_c\left[{\sigma^2}/({4\bar E_c^2})+ T\alpha^2 N/ \bar E_c\right]$.
We see that a thermodynamically infinitesimal broadening $\sigma$ 
of quantum levels generates a finite broadening  $\delta E\simeq 2\pi T\alpha^2 N$ due to the combined action of
temperature and disorder giving rise to a smooth $I$-$V$ characteristics.  This is in complete analogy to introducing a
finite width of  quantum levels of a given subsystem
of a larger thermodynamic system when
proving the equivalence of microcanonical, canonical, and grand canonical ensembles in quantum statistics.
The finite width appears due to interaction of this subsystem with the thermostat but disappears from the final expressions.
To conclude at this point, we have demonstrated  that at high temperatures, $T>{\bar E}_c$,
 the environmental spectrum is quasicontinuous and the system is in a conducting state. 
 At $T<{\bar E}_c$ the environmental spectrum retains its discrete character and transport is suppressed.  
 The localization-delocalization transition temperature is well defined by the condition
${\cal R}e=1$, where the first Landau-Hopf bifurcations appears.

The evolution of the low-voltage $I$-$V$ curves~\eqref{eq:Is_2n_gg_ap} upon going from high, 
$T> {\bar E}_c$, to low, $T<{\bar E}_c$,
temperatures is illustrated in Fig.~\ref{fig.IV}.
Panels a)-c) show first a practically smooth linear $I$-$V$ dependence at $T=10{\bar E}_c$ which then develops
pronounced modulations at resonance voltages upon cooling,
and finally at $T={\bar E}_c$ transforms into a dense set of resonant spikes.
Panels d) and e) show that at $T<{\bar E}_c$ the $I_s(V)$ dependence is a 
palisade of distinct resonant voltages apparently having a hierarchical structure,
reflecting the hierarchical glassy-like structure of the environmental spectrum.

Having derived the $I$-$V$ curves at various temperatures, one can determine the temperature dependence of 
the linear resistance $R(T)$ at small $V\to 0$ voltages by integration of the current over a small voltage interval.  
At high temperatures, $T>\bar{E}_c$, the resistance is determined by the
linear fit of the $I$-$V$ curve at low voltages. 
Below $T=\bar{E}_c$ the fractalized hierarchical peak structure of the $I$-$V$ dependence develops 
and the notion of a smooth $I$-$V$ characteristics ceases to exist. 
In order to extract a meaningful resistance value at 
low temperatures, one finds the electric power integrals over a small voltage interval 
near $V=0$ of the $I-V$ characteristics and determines the resistance from the relation $V_m^2/R=\int_0^{V_m}I(V)dV$.
The value $V_m$ is small as compared to $\bar E_c/e$, but still large enough to capture several peaks.
 Since one can expect that the peaks are rather narrow in experiments, the current baseline, 
 i.e. the straight current line limiting the current from below, 
 gives a good approximation for the resistance as well.

\begin{figure}[thb]
    \includegraphics[width=0.8\columnwidth]{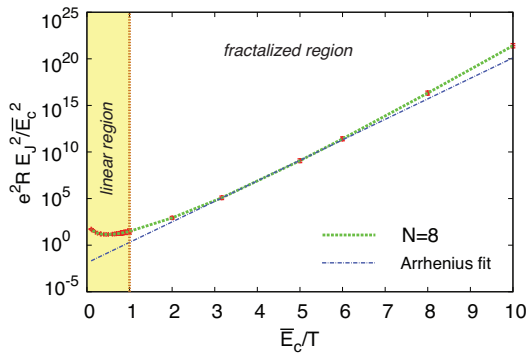}
  \caption{{\bf Superactivation} Dependence of the resistance $R$ on inverse temperature using the same parameters as in Fig.~\ref{fig.IV}. At temperatures just above $\bar{E}_c$ Arrhenius behavior is found, while at very high temperatures the resistance goes over in a power-law. The dotted line separated the linear behaviour of the current (delocalized region) and the fractalized behaviour (localized state at low temperatures).  Since the I-V characteristics becomes fractalized at low temperatures, the resistance is extracted by integration of the current at low voltages.
   }\label{fig.RT}
\end{figure}

Asymptotic expressions for the low- and high-temperature limits of the resistance 
can be written as
\begin{equation}\label{eq.asymR}
		R_{\rm asym}\propto\exp\left(\frac{N{\bar E}_c}{2T}+\delta_T\log(T/\bar{E}_c)\right)\,,
\end{equation}
with $\delta_T=-(N-3)/2$ for $T/\bar{E}_c< 1$ and $\delta_T=3/2$ for $T/\bar{E}_c> 1$ 
is derived from the exact Eq.~\eqref{eq:Is_2n_gg_ap}.
The numerically computed plot of $\ln R(T)$ vs. $1/T$ is shown in Fig.~\ref{fig.RT}.
The high temperature limit suggest that Arrhenius behaviour should be found for temperatures  
$\bar{E}_c<T<(N/3-1)\bar{E}_cW^{-1}(N/3-1)$
(for $N\geq 3$, where $W(x)$ is the Lambert $W$-function, defined as the inverse function of 
$f(W)=We^W$, with $W^{-1}(5/3)\approx 1.3$ for $N=8$).
Whereas in the low temperature limit, $T<\bar E_c$, the deviation from the Arrhenius exponent 
grows logarithmically with lowering the temperature. An intermediate Arrhenius behaviour is indicated 
by a linear fit in Fig.~\ref{fig.RT}.
Note, that the `activation' energy $E_a\simeq N\bar E_c$; this reflects the \textit{macroscopic} 
character of the Coulomb blockade effect
that governs the insulating behaviour of the JJA: an extra Cooper pair placed in a granule polarizes the whole system, and
since the Coulomb energy of two interacting charges grows linearly with the separation between the charges in 1D,
the activation Coulomb barrier is indeed $\propto N\bar E_c$~\cite{FVB,VinNature}.
At large temperatures one observes an upturn in the $\ln R(1/T)$ dependence (see Fig.~\ref{fig.RT}) in a nice accordance
with the experimental observations in TiN films~\cite{Kalok2010}.

\section{Discussion and Conclusions}

\begin{figure}[htb]
    \includegraphics[width=0.8\columnwidth]{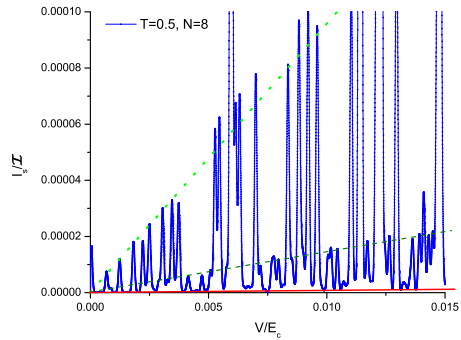}
  \caption{{\bf Resistance at small T} Low voltage behavior of the current at low temperatures (here $T=0.5\bar{E}_c$) for $N=8$, $\sigma=10^{-4}\bar E_c$, and $\alpha=0.056$. This plot shows the hierarchy of current peaks (green dotted and dashed lines) and the current baseline (red solid line). The current baseline (which limits $I_s/{\cal I}$ from below) can be used to extract an approximative value for the resistance which is observed in experiments at low temperatures. However, a more accurate value is obtained by calculating the electric power (see text).
   }\label{fig.lowT}
\end{figure}

The derived $I$-$V$ characteristic for a 1D system of Coulomb-interacting bosons in the presence
of relatively strong disorder represents an analytical description of a finite temperature phase transition between two
distinct dynamic states with different conducting and energy relaxation properties.
In the energy space, this
transition manifests itself as a transition between a `laminar' and `turbulent' spectral flow of the environment energy levels
occurring at the critical Reynolds number ${\cal R}e^*=(T/\bar E_c)=1$.  The evolution of the $I$-$V$ behaviour
shown in Fig.~\ref{fig.IV} indicates that for $T<\bar E_c$ the distribution of 
the resonant voltages, i.e. the distribution of characteristic energies
of the environment acquires a hierarchical structure (see Fig.~\ref{fig.lowT}).  
We thus can conjecture that at the transition temperature ($T^*\simeq \bar E_c$)
the dipole environment freezes into a glassy state where dipoles get localized, 
their local spectrum acquires a gap $\bar E_c$, and they become inefficient
to mediate energy relaxation of tunnelling carriers.
This suggests the one can relate the formation of a glassy state -- which basically can be viewed as a transition between
the two dynamic states, localized and conducting --
 on a generic level with the onset of turbulence in the spectral flow of the
respective energy spectrum of the system involved.

The obtained results are valid, strictly speaking, at low voltages $eV<{\bar E}_c$, since the use of an equilibrium distribution function presumes
that the time between the charge tunnelling events has to be greater than the relaxation time and thus the voltage has to be sufficiently low.
The observation of a dc tunnelling transport in large JJAs described above is feasible in experiments with artificially manufactured
1D systems with large enough ratio of the capacitance to the ground of the superconducting grains with respect to the average capacitance of the junction ($C/C_0$) such that the electrostatic screening length $\Lambda$ exceeded the perimeter of the system, $L$.
Even in the case where $\Lambda<L$ one can expect significant suppression of the conductivity 
in the low-temperature region, although the
characteristic energy scale will reduce to ${\bar E}_c\tilde N$ in that case, where $\tilde N=N\Lambda/L$, effectively replacing
the number of junctions $N$ by $\tilde N$.  
Our findings are in accord with early results by~\cite{Haviland} where the suppression of the conductivity
at low-temperatures and the corresponding voltage threshold behaviour was observed in the long 1D JJA arrays.  Furthermore,
our analytical results corroborate the findings of~\cite{FVB,VinNature} about the formation of the superinsulating state 
with the drastically suppressed conductivity at $T<E_{\mathrm c}$.

The results obtained for a JJA ring can be generalized onto 2D JJAs, since the resulting
transport across 2D JJAs can be presented as a superposition of  ring-like currents over overlapping closed loops spanning the 2D system.  This opens a remarkable
opportunity for constructing a comprehensive quantitative description of the glass formation and localization transition in higher dimensions
which will be discussed elsewhere.

\acknowledgments

We are grateful to M. Fistul who made an important contribution at the early stage of this work.
We like to thank B. Altshuler, I. Aleiner, and Yu. Galperin for illuminating discussions.
This work was supported  by the U.S. Department of Energy Office of Science through the contract DE-AC02-06CH11357, 
and partially by the ANR grant 09-BLAN-0097-01/2, 
by the Program ``Quantum Physics of Condensed Matter'' 
of the Russian Academy of Sciences, by the Russian Foundation for Basic Research 
(Grant No. 09-02-01205).


\begin{thebibliography}{99}
\bibitem{noteLH}
Note that Landau-Hopf scenario does not actually describe the turbulence in real liquids since it misses
the important nonlinear effects relevant or fluid dynamics.  This however is irrelevant for our purposes since
what is essential for us is the self-consistent mathematical structure of the Landau-Hopf theory and the prediction that 
holds (within the framework of the model) about the onset of stochastic dynamics that occurs upon exceeding 
the critical Reynolds number.
\bibitem{relax}
We consider the situation where the disorder dispersion $\alpha\bar{E}_c\gg E_J$, such that the conducting band for Cooper pairs does not form.  
This differentiates our system from the model of Ref.~\cite{SEA2009}.
\end{thebibliography}
\end{document}